%% file: ParametricColoringarXiv.tex
\pdfoutput=1

\documentclass[final,3p,times]{elsarticle}
\usepackage{graphicx}

\usepackage{amssymb}
\usepackage{amsmath}
\usepackage{amsthm}

\newtheorem{theorem}{Theorem}

\newtheorem{lemma}{Lemma}
\newtheorem{corollary}{Corollary}
\newtheorem{definition}{Definition}

\usepackage{algorithm}
\usepackage{algorithmic}
\usepackage{url}
\usepackage{array}
\usepackage{subfig}
\usepackage{stfloats}
\usepackage{booktabs}

\journal{Arxiv}
\begin{document}

\begin{frontmatter}



\title{Solving Hard Computational Problems Efficiently: Asymptotic Parametric Complexity 3-Coloring Algorithm}


\author{Jos\'e Antonio Mart\'in H.} \ead{jamartinh@fdi.ucm.es}
\address{Computer Architectures and Automation, Complutense University of Madrid, Spain}

\begin{abstract}
\input{abstract}

\end{abstract}


\end{frontmatter}


\input{body}

\clearpage

\section*{Tables}

\input{tables}

\clearpage
\bibliographystyle{elsarticle-num}
\bibliography{colorref}


\end{document}

%% file: abstract.tex
Many practical problems in almost all scientific and technological disciplines have been classified as computationally hard (NP-hard or even NP-complete).
In life sciences, combinatorial optimization problems frequently arise in molecular biology, e.g., genome sequencing; global alignment of multiple genomes; identifying siblings or discovery of dysregulated pathways.
In almost all of these problems, there is the need for proving a hypothesis about certain property of an object that can be present only when it adopts some particular admissible structure (an NP-certificate) or be absent (no admissible structure), however, none of the standard approaches can discard the hypothesis when no solution can be found, since none can provide a proof that there is no admissible structure.
This article presents an algorithm that introduces a novel type of solution method to ``efficiently'' solve the graph 3-coloring problem; an NP-complete problem. The proposed method provides certificates (proofs) in both cases: present or absent, so it is possible to accept or reject the hypothesis on the basis of a rigorous proof.
It provides exact solutions and is polynomial-time (i.e., efficient) however parametric. The only requirement is sufficient computational power, which is controlled by the parameter $\alpha\in\mathbb{N}$.
Nevertheless, here it is proved that the probability of requiring a value of $\alpha>k$ to obtain a solution for a random graph decreases exponentially: $P(\alpha>k) \leq 2^{-(k+1)}$, making tractable almost all problem instances.
Thorough experimental analyses were performed. The algorithm was tested on random graphs, planar graphs and 4-regular planar graphs. The obtained experimental results are in accordance with the theoretical expected results.

%% file: body.tex
%
%
%
%
%

\section*{Introduction}
\label{Introduction}
Graph Coloring is one of the oldest and among the most popular Constraint Satisfaction Problems (CSPs)~\cite{jones2008artificial}. The study of efficient CSP-solving algorithms is a central topic in Computer Science and Artificial Intelligence because of its wide applicability in many engineering projects, e.g., very-large-scale integration (VLSI) testing, planning and scheduling, timetabling, satellite range scheduling, register allocation, printed circuit testing, and frequency assignment~\cite{Wigderson1983,park1996,ramani2004}, as well as theoretical physical models, e.g., spin-glasses and the anti-ferromagnetic Potts model~\cite{zdeborova2007}.

Graph coloring has found application in life sciences as well, for instance, nucleic acid sequence design has been modeled as a graph coloring problem~\cite{Ingrid2005}, and in general, combinatorial optimization problems frequently arise in molecular biology: genome sequencing; global alignment of multiple genomes; identification of siblings, cousins, or second cousins through comparison of genomes; finding protein modules containing specified types of proteins; or the computational discovery of dysregulated pathways in human diseases are NP-hard or even NP-complete problems~\cite{pevzner1995open,karp2011heuristic}\footnote{The computational complexity terminology and concepts (e.g., P, NP, CoNP, NP-completeness, NP-Hardness, certificate/witness, and reductions) can be consulted in the recent books of Arora and Barak~\cite{Arora2009} and Goldreich~\cite{goldreich2008}.}.

The graph coloring problem involves assigning a number $c\in\{1,2,3,\ldots,k\} $ (i.e. a color) to each vertex of a graph such that neighboring vertices are assigned different colors. The problem of deciding whether a given graph can be colored with \emph{k} or less colors is called the \emph{k}-colorability problem.

For $k = 2$, the colorability problem can be solved efficiently. However, for $k \geq 3$, in general, there is no known efficient algorithm to determine whether the graph is \emph{k-colorable} and the problem is NP-complete~\cite{cook1971,karp1972reducibility,levin1973universal,GJ79}.

Direct methods for NP-complete problems require the determination of simple properties (e.g. triangle freeness) that impose necessary and sufficient conditions for determining the class of an instance (Yes/No). For the three-color-problem~\cite{steinberg1993state}, two examples are as follows: phase-transition studies~\cite{hogg1996phase,Culberson2001,mulet2002,Boettcher2004,zdeborova2007}, based on random graphs theory and sharp thresholds~\cite{erdos1960,erdos1973asymptotic}, and structural-combinatorial approaches based on certain specific parameters such as the existence or absence of particular cycle configurations~(see \cite{steinberg1993state,Borodin1996,Borodin2005,Wang2007,Borodin2009}). One of the foundational results of this approach is the Grotzsch's 3-color theorem~\cite{Thomassen1994}: the triangle-free planar graphs are 3-colorable.

On the other hand, given the intractability of NP-complete problems, research on approximation algorithms began early (e.g., \cite{Johnson1974,johnson1974worst,Garey1976nearoptimal,Wigderson1983,berger1990better,halldorsson1993still,blum1994new,Arora2006}). However, even for the approximate case graph coloring remains hard~\cite{karger1998approximate}. More strictly, it is NP-hard to even find a 4-coloring of a 3-chromatic graph~\cite{khanna2000hardness}.

A recently developed alternative approach to the classical worst-case computational complexity theory is parameterized complexity~\cite{downey1999parameterized,flum2006parameterized}. In parameterized complexity, apart from the problem instance itself, there is a parameter (usually an integer) that may be associated arbitrarily with the problem instance, allowing one to study a problem's complexity with respect to both the size of the input (as in classical computational complexity) and the provided parameter. One of the main virtues of the parametric complexity approach is the concept of fixed-parameter tractability. Here, instead of fixed-parameter complexity, we present an algorithm with \emph{asymptotic} parameter complexity. Two key differences with respect to the past graph coloring approaches are that it is (1) exact and (2) polynomial-time but parametric, while previous algorithms are either exact but low-exponential or polynomial-time but approximate.

A very easy but naive way of coloring a graph is simply sequentially assigning the first available color to each vertex. Such an algorithm is known as a greedy coloring algorithm. The base procedure for the proposed algorithm also follows a greedy approach (greedy contraction): sequentially contract two non-neighboring vertices until either a triangle (representing a legal 3-coloring) or a graph containing a 4-clique ($K_4$) subgraph (a non-3-colorable subgraph) is obtained. However, this greedy algorithm often fails when the graph is 3-colorable since some contractions will unavoidably lead to a $K_4$ subgraph. The key idea of the proposed algorithm is a way of determining which contractions will fail and avoid such a failure by adding a new edge \emph{uv} instead of doing a contraction \emph{G/uv} so that if the graph is 3-colorable, the greedy contraction algorithm will necessarily converge to a triangle, i.e., a legal 3-coloring and if it is non-3-colorable, a 3-uncolorability certificate will be obtained.

A 3-uncolorability certificate is defined as a sequence of ``unavoidable vertex contractions'' leading to a graph containing a $K_4$. A verifier can efficiently check whether every contraction is ``unavoidable''. The possibility of efficiently generating 3-uncolorability certificates is of theoretical interest since 3-colorability is NP-complete, and thus 3-uncolorablity is CoNP-complete. Previous works have studied short proof systems for CoNP-complete problems (e.g., \cite{Boppana1987, Fortnow1988}). In particular, for graph coloring, there is a recent work~\cite{Bes2005} that tries to find graph uncolorability proofs on the basis of consistency checks of CSPs. However, short 3-uncolorability certificates are not possible in general unless NP = CoNP (which remains unknown).

The proposed algorithm has several ``good'' (desired) key features:

\begin{enumerate}
 \item The proposed algorithm is exact and runs in polynomial-time; however, it is parametric. Its running time can be controlled (bounded) by means of a simple parameter ($\alpha$, the maximum recursion level) that determines the order of the bounding polynomial. Hence, its complexity is on-demand.
 \item If for a given $\alpha$ the algorithm is unable to find a certificate then it returns ``undetermined'' so that it can be re-run with a higher $\alpha$ until a solution is obtained, taking into account the computational resources available. Self-tuning the $\alpha$ parameter (by using the undetermined return value) gives the algorithm the capability of using only the required computational resources for a particular problem instance, e.g., for almost all planar graphs, it is sufficient to have a value of $\alpha = 0$ to obtain the right solution, thus obtaining an $O(n^2)$ algorithm for the 3-coloring problem in almost all planar graphs.
 \item It generates certificates for both Yes and No instances, i.e., either a legal 3-coloring or a 3-uncolorability certificate, and hence, it gives stand-alone indubitable results so that it is not necessary to trust neither the correctness of the algorithm itself nor the particular implementation used for recognizing whether the provided solution is correct, since the result can be efficiently verified using only the provided solution (the certificate or witness).
\end{enumerate}

Moreover, from the theoretical point of view, the most important result is the classification of all graphs by the number $\alpha(G)$: the minimum value of the parameter $\alpha$ required by the algorithm to obtain a certificate given a particular instance $G$ of the 3-coloring problem\footnote{A rigorous definition assuring that $\alpha(G)$ is well-defined is presented in the proof of the main theorem section.}. This results in important consequences and allows the development of a thorough analysis.

Since for each finite graph \emph{G} there is a corresponding $\alpha(G)\in \mathbb{N}$, the algorithm is polynomial-time, and its order depends on $\alpha$,
\begin{align}
\bigcap_{\alpha=0}^{\infty} \text{NP} \setminus \text{P} = \varnothing,
\end{align}
since P also depends on $\alpha$. However, for both practical and theoretical results, the most important thing to be determined is the ``speed'' of this convergence. Here, as the main theoretical result, it is proved, the non-trivial and highly significant fact, that the probability of requiring a value of $\alpha>k$ for obtaining a solution for a random graph decreases exponentially as a function of $k$. This result is formalized as Theorem~\ref{thm:alpha:distribution} given below:

\begin{theorem}\label{thm:alpha:distribution} Let $G$ be a random graph; if $P(\alpha = k )$ is the probability that $\alpha(G) = k$ and $P(\alpha > k )$ is the probability that $\alpha(G) > k$ then for all $k\in\mathbb{N}$:
\begin{align}
P(\alpha = k )           &\geq P( \alpha > k),\\
P\left(\alpha > k\right) &\leq \quad 2^{-(k +1)}.
\end{align}
\end{theorem}

In the experimental part of this article, the algorithm was thoroughly evaluated using significant samples pertaining to three different graph distributions:
\begin{enumerate}
  \item Random planar graphs~\cite{denise1996,bodirsky2003}.
  \item Random 4-regular planar graphs~\cite{manca1979,lehel1981,broersma1993}.
  \item Erd\H{o}s-R\H{e}nyi connected random graphs~\cite{erdos1960}.
\end{enumerate}

An interesting experimental finding is that in all the test cases for planar graphs, it was found that fixing the maximum recursion level to $\alpha = 1$ was sufficient to obtain a solution, i.e., exact and efficiently verifiable results were obtained using a polynomial algorithm. This was also the case for 4-regular planar graphs. Furthermore, in the general (random graphs) 3-coloring case experiments, it has been observed that the distribution of $\alpha(G)$ conforms to the theoretical decreasing pattern: the majority of graphs are in $\alpha(G) = 0$ or $\alpha(G) = 1$, some in $\alpha(G) = 2$, a few in $\alpha(G) = 3$, very few in $\alpha(G) = 4$, and so on. Indeed, it was not possible to obtain a graph with $\alpha(G) > 4$ in the sampled random graphs.

\subsection*{Practical applicability}

The most common methods for dealing with a NP-complete problem when solving a big (intractable by brute force) real problem are: heuristic algorithms, approximate algorithms, randomized algorithms, and fixed-parameter tractability. The presented algorithm introduces a novel type of solution method and presents many new features that are usually absent from the previous approaches.

Suppose that a very critical hypothesis about certain phenomenon needs to be proved in a molecular biology study and that such a hypothesis depends on testing a property of an object that can be present only when it adopts some particular admissible structure (an NP-certificate) or be absent (no admissible structure for this property) and the problem is not fixed-parameter tractable. Then,

\begin{enumerate}
  \item None of the standard approaches can discard the hypothesis when no solution is found, since none will give a proof that the problem has no solution, i.e., a proof that there is no admissible structure for this property.
  \item Even when a solution exists, heuristic as well as randomized algorithms do not guarantee finding a solution even with extraordinary computational power.
  \item Approximate algorithms can give, if lucky, an estimated probability of the existence of such a property, possibly including an approximate admissible structure that is only probably correct.
\end{enumerate}

However, the proposed method solves the problem by providing certificates (proofs) in both cases: present or absent; hence, one can accept or reject the hypothesis on the basis of a rigorous proof that will be independent of the algorithm itself and the implementation used. Moreover, the proposed method assures an exact solution. The only requirement is sufficient computational power (as in brute force methods). However, we have proven that the amount of required computational resources, i.e., the complexity of a problem for the proposed method, is distributed negative exponentially with respect to the problem complexity; hence, the harder a problem is, the lower is its probability of appearance. The exponential reduction in unsolvable instances makes any investment in computing power profitable.

\subsection*{Basic terminology}
This article follows the standard graph theory terminology; for general terms and notation, the book of Jensen and Toft~\cite{JT95} and the recent book on chromatic graph theory by Chartrand and Zhang~\cite{Chartrand2008} should be consulted. However, some special terms and particular notations are defined below. Unless we state otherwise, all graphs in this work are connected and simple (are finite and have no loops or parallel edges).

The term \emph{random graph} $G_{n,m}$ refers to a graph chosen at random (with equal probability) from among all possible graphs with $n$ vertices and $m$ edges, as defined by Erd\H{o}s-R\H{e}nyi~\cite{erdos1960}.

We refer to $u,v$ as a \emph{planar preserving edge} if $uv$ is not an edge of $G$ and $G + uv$ remains planar. A \emph{vertex contraction}, also called vertex identification or vertex merging, is denoted by $G/uv$. Vertex or edge additions and deletions are denoted as follows: $G + u$ or $G + uv$ and $G - u$ or $G - uv$, respectively. A vertex ordering of a graph $G = (V, E)$ is a bijection $\pi\; :\; V \; \rightarrow \; \{1, 2, . . ., |V |\}$, and thus, a set of \emph{n} vertices can be ordered in $n!$ different ways.

\begin{figure}[tb]
\begin{center}
\includegraphics[width=6in]{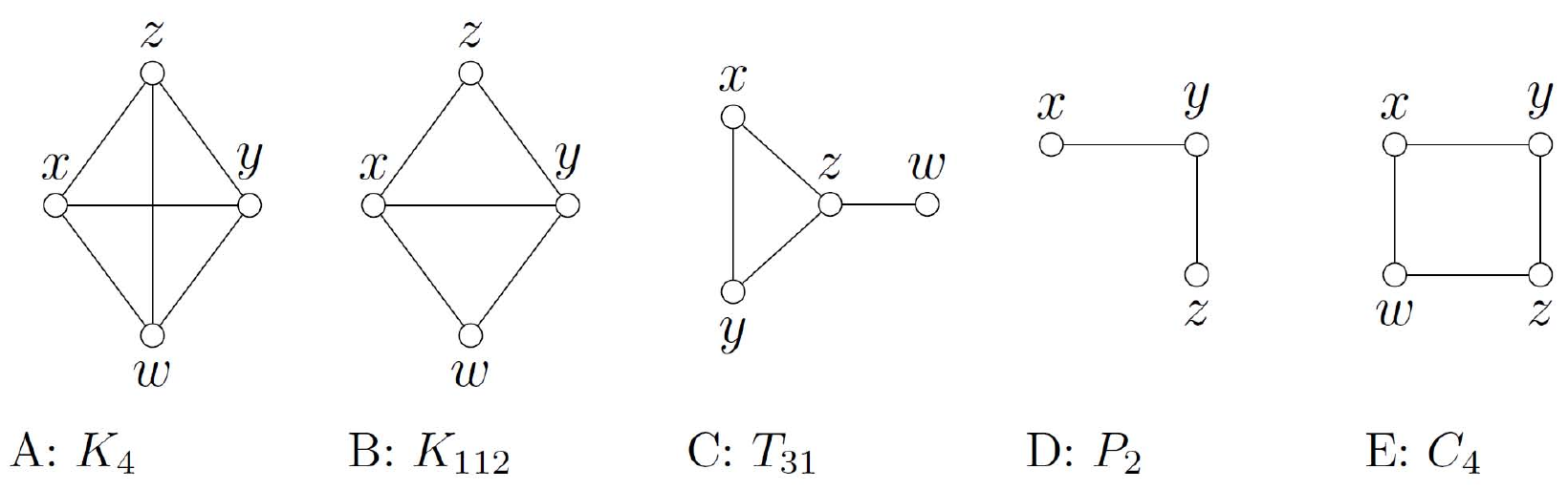}
\end{center}
\caption{
{\bf Specially named graphs used in the article.} (A) Complete graph $K_4$. (B) Complete 3-partite graph $K_{112}$. (C) Tadpole graph $T_{31}$, which imposes a binary constraint on 3-coloring: either $G/xw$ or $G/yw$. (D) Path of length two $P_2$. (E) $C_4$ graph that imposes a binary constraint on 3-coloring: either $G/xz$ or $G/yw$.}
\label{graph:graphs}
\end{figure}

The specially named graphs used in the article are as follows (c.f. figure~\ref{graph:graphs}): complete graph $K_{4}$, diamond graph, complete 3-partite graph $K_{112}$ (a $K_{4}$ minus one vertex), \emph{tadpole} graph $T_{31}$, \emph{triangle} graph $T_3$, path graph $P$ of length two $P_{2}$, and square graph $C_4$.

A \emph{certificate}~\cite{Arora2009} (or witness~\cite{goldreich2008}) is an efficiently verifiable proof of the correctness of an answer for some given decision problem. For instance, given a graph $G$, a legal 3-coloring of $G$ or a short proof that $G$ is not 3-colorable
are certificates for the 3-colorability problem.

\section*{Materials and Methods}

\subsection*{Definition of the Algorithm}
\label{sec:definitions}
\begin{definition}
Given a graph $G$ and a 3-colorable subgraph $H$ of $G$, there is an \emph{unavoidable vertex contraction}\footnote{Many works define the same relation between such a pair of vertices, e.g., ``two nodes $u,v$ of a given graph $G$ are 3-color bound (or simply bound) if $u$ and $v$ must be assigned the same color in any 3-coloration of $G$''~\cite{Stockmeyer1973}. ``Two vertices of a graph are said to be 3-chromatically connected if they are assigned the same color in any 3-coloring of the graph''~\cite{steinberg1993state}. Culberson and Gent~\cite{Culberson2001} also use the term ``frozen pair''. Moreover, the same relation for $u,v$ is called ``implicit identity'' by the current author, who presents~\cite{martinh2011} a thorough study of this subject for the general case (k-chromatic and non-planar graphs).} $u,v \in V(G),\; u,v \notin E(G)$ if the addition of the new edge $uv$ to $H$ makes $H$ not 3-colorable.
\end{definition}

\begin{definition}\label{def:uncolor}
Given a non-3-colorable input graph $G$, a \emph{3}-\emph{uncolorability} \emph{certificate} $\mathcal{W}$ is a description of a (possibly empty) sequence of unavoidable vertex contractions, $G/uv$, leading to a graph containing $K_4$, such that either of the following two cases apply:
\begin{enumerate}
  \item $u,v$ are the non-complete vertices of a complete $3$-partite $K_{112}$ diamond subgraph of $G$ or;
  \item a nested 3-uncolorability certificate for the graph $G+uv$ is provided.
\end{enumerate}
\end{definition}

Hence, in order to design an algorithm for obtaining a 3-uncolorability certificate, a method for obtaining such nested certificates should be provided. The proposed algorithm is recursive and uses a parameter $\alpha$ to limit the recursion depth. A very simple sketch of this algorithm is as follows:
\begin{description}
  \item[Algorithm: 1] is-3-colorable$(G,\alpha)$:
  \item[1] Contract every $u,v$ of a diamond subgraph until no other diamond subgraph exists or until the graph becomes the $K_3$ or it contains a $K_4$ subgraph.
  \item[2] If the graph becomes the $K_3$ graph then return the current contraction sequence (i.e., a legal 3-coloring).
  \item[3] If $K_4$ is found then return the current contraction sequence (i.e., a 3-uncolorability certificate).
  \item[4] If the current recursion level $\alpha = 0$ then return ``undetermined for the current value of $\alpha$."
  \item[5] For each non-edge $u,v$,
  \begin{description}
  \item[5.1] If not is-3-colorable$(G+uv,\alpha-1)$ then
  \begin{description}
  \item[5.1.1] Contract $u,v$.
  \item[5.1.2] Append the nested certificate for $G + uv$.
  \item[5.1.1] Break and continue at step 1.
  \end{description}
  \end{description}
  \item[6] Return ``undetermined for the current value of $\alpha$."
  \item[END.]
\end{description}

Now, let us define a greedy 3-coloring algorithm that will serve as the baseline for the derivation of the proposed coloring algorithm.

\begin{definition}
The $g_3(G)$ algorithm is a ``greedy-contraction'' 3-coloring algorithm that sequentially, and at each step, selects two non-adjacent vertices $x$ and $y$ of a graph $G$ and contracts them to obtain the graph $G/xy$, while maintaining a list $S$ of the vertices that have been contracted thus far so that if the resulting graph is a triangle (or even a $K_2$) and $S$ contains at most three independent sets, these are three (or less) color classes of $G$ and hence, a legal 3-coloring of $G$.
\end{definition}

The justification for using such a simple approach in combination with a more sophisticated way of detecting (and avoiding) vertex contraction that unavoidably leads to an non-3-colorable graph is derived from the following lemma (lemma~\ref{lemma:greedy:coloring}):

\begin{lemma}
\label{lemma:greedy:coloring}
Given an exact algorithm $\mathcal{W}(G)_0$ of complexity $O(n^k)$ to obtain a 3-uncolorability certificate for any non-3-colorable graph, there is an exact algorithm $\mathcal{W}(G)_1$ of complexity $O(n^{k + 1})$ to obtain a 3-coloring of any 3-colorable graph.
\begin{proof}
Assume $\mathcal{W}(G)_0$ exists. Then, given a 3-colorable graph $G$, apply the greedy $g_3(G)$ algorithm but avoiding the contraction of every $\{x,y\}$ such that $G/xy$ is not 3-colorable, which can be determined in $O(n^k)$ by $\mathcal{W}(G/xy)_0$. Since $G$ is 3-colorable, it will converge (at most) to a triangle graph. Since $g_3(G)$ is of complexity $O(n)$ (at every step, at least one vertex will get colored), we obtain a 3-coloring of $G$ in $O(n)O(n^k) = O(n^{k + 1})$
\end{proof}
\end{lemma}

\begin{corollary} Hence, if $\mathcal{W}(G,\alpha)_0$ is an exact parametric algorithm, of complexity $O(n^{f(\alpha)})$, to obtain a 3-uncolorability certificate for any non-3-colorable graph, there is an exact parametric algorithm $\mathcal{W}(G,\alpha)_1$ of complexity $O(n^{f(\alpha) + 1})$ to obtain a 3-coloring of any 3-colorable graph.
\end{corollary}

Hence, the basic complete coloring algorithm can be described as follows:
\begin{description}
  \item[Algorithm: 2] general-3-COL$(G,\alpha)$:
  \item[1] If not is-3-colorable$(G,\alpha)$ then return 0 and the 3-uncolorability certificate.
  \item[2] While $G$ has more than three vertices,
  \begin{description}
  \item[2.1] Select two non-neighboring vertices $u,v$.
  \item[2.2] If not is-3-colorable$(G/uv,\alpha)$ then $G \leftarrow G + uv$.
  \item[2.3] Else, $G\leftarrow G/uv$.
  \item[2.4] If $K_4 \in G$, return $\infty, \varnothing$.
  \end{description}
  \item[3] Return 1 and a legal 3-coloring as the list of contracted vertices.
  \item[END.]
\end{description}

Finally, an automated algorithm can be developed to eliminate the need for specifying the $\alpha$ parameter.
\begin{description}
  \item[Algorithm: 3] BFS\_3COL(\emph{G}):
  \item[1] For $\alpha = 0 \; {to} \; \infty$
  \begin{description}
  \item[1.1] If general-3-COL$(G, \alpha) = 0$, return a 3-uncolorability certificate.
  \item[1.2] If general-3-COL$(G, \alpha) = 1$, return a legal 3-coloring.
  \end{description}
  \item[END.]
\end{description}

\subsection*{Some advanced improvements and special case handling}

The algorithm is divided into two parts: the decision problem (\texttt{is-3-colorable}) and the coloring algorithms (\texttt{general-3COL}). There are two versions of each of these algorithms: one for planar graphs and the other for non-planar graphs. First, the algorithm for the planar graphs case is described, which is better for understanding the key idea behind the algorithms. Then, this description is generalized for the non-planar graph case.

\subsubsection*{Specialization for planar graphs}

The development of a special algorithm for planar graphs has two main advantages:
\begin{enumerate}
  \item To take advantage of some special structural constraints of planar graphs (e.g., Gr\H{o}tzsch's like theorems) that aid the development of more efficient algorithms.
  \item To formalize an algorithm for planar graphs that preserves planarity at each step, allowing the development of theoretical studies on the the class of planar graphs, e.g., inductive proofs and structure-based proofs.
\end{enumerate}

Now, lets us consider the (\texttt{is-3-colorable}) routine. According to Grotzsch's 3-color theorem~\cite{Thomassen1994} (triangle-free planar graphs are 3-colorable), every non-3-colorable planar graph should have \{\emph{x,y,z,w}\}-tadpole $T_{31}$ subgraphs (cf. Figure~\ref{graph:graphs}B). The key idea is that $T_{31}$ subgraphs impose binary constraints, i.e., either $\{x,w\}$ or $\{y,w\}$ must be contracted since $T_{31} + xw + yw$ is a $K_4$ (the same is true for square graphs). Thus, there is no need for Step 5 of Algorithm 1 to check every non-edge but just every $T_{31}$ subgraph. Thus, the routine can be performed for each $T_{31}$ by contracting $G/yw$ whenever $G/xw$ is not 3-colorable, i.e., when \emph{y, w} is a unavoidable vertex contraction, as shown in the next algorithm:

\begin{description}
  \item[Algorithm: 4] is-3-colorable-planar$(G,\alpha)$:
  \item[1] Contract every $u,v$ of a diamond subgraph until no other diamond subgraph exists or until the graph becomes the $K_3$ or it contains a $K_4$ subgraph.
  \item[2] If the graph becomes the $K_3$ graph then return the current contraction sequence (i.e., a legal 3-coloring).
  \item[3] If $K_4$ is found then return the current contraction sequence (i.e., a 3-uncolorability certificate).
  \item[4] If the current recursion level $\alpha = 0$ then return ``undetermined for the current value of $\alpha$."
  \item[5] For each \{\emph{x,y,z,w}\}-tadpole $T_{31}$ subgraph,
  \begin{description}
  \item[5.1] If not is-3-colorable$(G/xw,\alpha-1)$ then
  \begin{description}
  \item[5.1.1] Contract $yw$.
  \item[5.1.2] Append the nested certificate for $G + yw$.
  \item[5.1.1] Break and continue at Step 1.
  \end{description}
  \end{description}
  \item[6] Return ``undetermined for the current value of $\alpha$."
  \item[END.]
\end{description}

Now, let us show a planarity preserving coloring algorithm for planar graphs. The idea involves reducing $G$ to a planar triangulation by means of the addition/contraction of the planar preserving edges of the planar graph $G$. At the end, if the triangulation has all degrees even, it is 3-colorable~\cite{heawood1898four,JT95} and finding a legal 3-coloring is linear-time. Otherwise, the algorithm returns ``undetermined,'' meaning that $\alpha$ was not sufficient for obtaining a certificate for the input graph. The specialized coloring routine is described next.

\begin{description}
  \item[Algorithm: 5] general-3-COL-planar$(G,\alpha)$:
  \item[1] If not is-3-colorable-planar$(G,\alpha)$ then return 0 and the 3-uncolorability certificate.
  \item[2] While $G$ is not a planar triangulation,
  \begin{description}
  \item[2.1] Select a planar preserving edge \emph{u, v}.
  \item[2.2] If not is-3-colorable-planar$(G/uv,\alpha)$ then $G \leftarrow G + uv$.
  \item[2.3] Else, $G\leftarrow G/uv$.
  \item[2.4] If $K_4 \in G$, return $\infty, \varnothing$.
  \end{description}
  \item[3] If triangulation $G$ has an odd vertex, return $\infty, \varnothing$
  \item[4] Return 1 and a legal 3-coloring of $G$ in linear time.
  \item[END.]
\end{description}
As can be seen, the graph remains planar at each step, making valid any assumption or structural property of planar graphs at each iteration.

\subsubsection*{A slight improvement of the worst and expected cases in non-planar graphs}

For non-planar graphs, a slight modification can be made to improve the worst and the expected case running time of the algorithm. The key idea in this case is to build a complete vertex, i.e., a vertex joined to all the remaining vertices of the graph so that for testing 3-colorability it is sufficient to test 2-colorability of the neighborhood, which can be done in linear time.

\begin{description}
  \item[Algorithm: 6] general-3-COL$(G,\alpha)$:
  \item[1] If not is-3-colorable$(G,\alpha)$ then return 0 and the 3-uncolorability certificate.
  \item[2] Let $u$ be the vertex with the highest degree of $G$.
  \item[3] While $u$ is not a complete vertex
  \begin{description}
  \item[3.1] Select a non-neighbor $v$ of $u$ such that the common neighborhood ($N(u)\cap N(v)$) is minimized.
  \item[3.2] If not is-3-colorable$(G/uv,\alpha)$ then $G \leftarrow G + uv$.
  \item[3.3] Else, $G \leftarrow G/uv$.
  \item[3.4] If $K_4 \in G$, return $\infty, \varnothing$.
  \item[3.5] If $N(u)$ is not bipartite, return $\infty, \varnothing$.
  \end{description}
  \item[4] If $N(u)$ is not bipartite, return $\infty, \varnothing$.
  \item[5] Return 1 and a legal 3-coloring as the list of contracted vertices.
  \item[END.]
\end{description}

\subsection*{Proof of the Main Theorem}
To formalize the analysis of the algorithm, let us define the following two algorithm specifications:

\begin{definition}
The $\mathcal{A}(G,\alpha)$ is a parametric-complexity algorithm that computes a function that assigns to a given input graph $G$ just one of three possible values: $0, 1$, or $\infty$, when $G$ is, respectively, non-3-colorable, 3-colorable or the algorithm was unable to find a solution for the given value of the $\alpha$ parameter:
\end{definition}

\begin{equation}
\label{eq:algorithm}
\mathcal{A}(G,\alpha)
\left\{
  \begin{array}{ll}
    0, & \hbox{No-instance: $G$ is not 3-colorable;} \\
    1, & \hbox{Yes-instance: $G$ is 3-colorable;} \\
    \infty, & \hbox{undetermined for the given} \alpha
  \end{array}
\right.
\end{equation}

\begin{definition}
The $\mathcal{W}(G,\alpha)$ is a parametric-complexity algorithm that computes a function that assigns to a given input graph $G$ just one of three possible values: a 3-uncolorability certificate, a legal 3-coloring, or a null value, when $\mathcal{A}(G,\alpha)$ is, respectively, $0,1$, or $\infty$.
\begin{equation}
\label{eq:algorithm}
\mathcal{W}(G,\alpha)
\left\{
  \begin{array}{ll}
   \hbox{a 3-uncolorability certificate}, & \mathcal{A}(G,\alpha) = 0;  \\
   \hbox{a legal 3-coloring},         & \mathcal{A}(G,\alpha) = 1;  \\
   \varnothing,                       & \mathcal{A}(G,\alpha) =  \infty
  \end{array}
\right.
\end{equation}
\end{definition}

Since the proposed algorithm is a greedy algorithm, it is affected as the other greedy sequential coloring algorithms, by the initial vertex ordering; hence, it is not possible to define a function $\alpha(G)$ simply as the minimum $k \in \mathbb{N}$ required to obtain a certificate for a particular graph $G$ without considering the vertex ordering; for instance, for any 3-colorable graph, it can be shown at least two different vertex orderings $V^1,V^2$ such that for $V^1$, a solution can be found for a value of $\alpha(G) = 0$, while for $V^2$, a value of $\alpha(G )> 0$ is required. Thus, a solution is to define the function $\alpha(G)$ on the basis of the worst-case vertex ordering, and therefore, $\alpha(G)$ will imply a computational complexity measure.

\begin{definition}
Given a graph $G$, the integer $\alpha(G)$ is
\begin{description}
  \item[For $G$ non-3-colorable:] The minimum $k \in \mathbb{N}$ required to obtain a 3-uncolorability certificate, assuming that the ordering of the vertices is the worst case for the is-3-colorable$(G,k)$ algorithm.
  \item[For $G$ 3-colorable:] The value $\alpha(H)$ of the non-3-colorable graph $H = G/uv$ where $\alpha(H)$ is the maximum over all $H = G/uv$ required to obtain a solution for $G$, assuming that the ordering of the vertices is the worst case for the general-3-COL$(G,k)$ algorithm.
\end{description}
\begin{equation}
    \alpha(G) \; =\; \max \big( \min(k) \; s.t. \; \mathcal{W}(G,k) \neq \varnothing \; \forall \pi \; \big).
    \end{equation}
\end{definition}

Now, let us divide the proof of Theorem~\ref{thm:alpha:distribution} into two cases:
\begin{enumerate}
  \item For non-3-colorable graphs, i.e., $\chi(G)\geq 4$.
  \item For 3-colorable graphs, i.e., $\chi(G) \leq 3$.
\end{enumerate}

First, it is proved that for non-3-colorable graphs the cardinality of the set $\mathbb{A}$ of graphs with $\alpha(G) = k$ is greater than the set $\mathbb{B}$ of graphs with $\alpha(G) > k$.

\begin{lemma} Let $\mathbb{G}^*$ be the set of all graphs and assume (with no loss of generality) that in particular $\mathbb{G}^*$ is defined for a maximum number of vertices or edges that exhausts the representation limit of any computational device, i.e., $\mathbb{G}^*$ is finite. Let $\mathbb{H}$, $\mathbb{A}$, and $\mathbb{B}$ be the sets:
\label{lemma:cardinality}
\begin{align}
\mathbb{H} &= \left\{G \in \mathbb{G}^* \;|\; \chi(G)\geq 4 \right\}, \\
\mathbb{A} &= \left\{G \in \mathbb{H} \;|\; \alpha(G) = k \right\}, \\
\mathbb{B} &= \left\{G \in \mathbb{H} \;|\; \alpha(G) > k \right\},
\end{align}
then
\begin{equation}
\left|\mathbb{B}| < |\mathbb{A} \right|.
\end{equation}
\begin{proof}
Since $\mathbb{H}$ is the set of non-3-colorable graphs, for every graph in $\mathbb{B}$, there is at least a graph in $\mathbb{A}$: simply take any graph in $\mathbb{B}$ not in $\mathbb{A}$ and join it to the smallest graph in $\mathbb{A}$; the resulting graph is in $\mathbb{A}$ since a 3-uncolorability certificate can be found with $\alpha(G) = k$. Moreover, no graph of $\mathbb{A}$ is a subgraph of any graph in $\mathbb{B}$. Hence, the cardinality of $\mathbb{B}$ is strictly less than the cardinality of $\mathbb{A}$.
\end{proof}
\end{lemma}

Therefore, case 1 is proved since $\mathbb{A}$ and $\mathbb{B}$ are finite sets and a uniform probability distribution over $\mathbb{C} =\mathbb{A}\cup \mathbb{B}$ is well defined. Hence, $P(\alpha = k)\geq P(\alpha > k)$ holds for non-3-colorable graphs.

Case 2 can be reduced to case 1 since for 3-colorable graphs $P(\alpha = k), \geq P( \alpha > k)$ reduces to $P(\alpha(H) = k ) \geq P( \alpha(H) > k)$ for the worst-case non-3-colorable graph $H = G/uv$ by the definition of $\alpha(G)$ for 3-colorable graphs.

Therefore, on the basis of the lemma~\ref{lemma:cardinality}, Theorem~\ref{thm:alpha:distribution} holds for all $k$ since $\left|\mathbb{B}| < |\mathbb{A}\right|$ implies that $P(\alpha = k) \geq P(\alpha > k)$, and hence, $P\left(\alpha > k\right) \leq 2^{-(k + 1)}$. This completes the proof of the main theorem.\qed

\subsection*{Runtime analysis of the algorithm}
The average-case complexity, worst-case complexity, and experimental performance of the algorithm are analyzed. The average-case analysis is informally presented as a mean of establishing the theoretically expected behavior over different kinds of instances. The worst-case analysis establishes the order (Big $O$) of the algorithm. Finally, the experimental analysis confronts the algorithm with samples from a series of graph distributions to study its performance and contrast it with the theoretical results.

\subsubsection*{Average-case (expected) complexity}
Table~\ref{table:analysis1} shows the average case (expected) performance of the algorithm with respect to the type of the instance (Yes/No) and the density of the graph, i.e., above/below the phase transition threshold.

In all the cases (except at the phase transition threshold), there is a high probability of a short running time. A priori, it may look that the worst case should occur on the sparse non-3-colorable graphs. This observation is based on the fact that for this class of graphs, it is more complex to obtain a $K_4$ by random edge additions and vertex contractions; nevertheless, some restrictions apply. Since the proportion of non-3-colorable graphs decreases fast below the threshold and almost all non-colorable graphs contain a $K_4$, the probability of obtaining a $K_4$-free non-3-colorable graph below the threshold is very small. Moreover, it is known that vertex 3-colorability of a graph with maximum vertex degree three can be determined in polynomial-time~\cite{steinberg1993state}. Further, every vertex of maximum degree two can be removed from the graph without affecting the 3-colorability; thus, non-3-colorable sparse graphs are very rare below $c \lesssim 4$ (this follows from the sharp-thresholds theory).

Hence, in almost all cases, a short running time is expected. By short, I mean significantly shorter than the worst-case upper bound.

\subsubsection*{Worst-case complexity}

To determine the computational complexity ($\mathrm{g}$) of the entire algorithm, we will start by analyzing the algorithm from the \texttt{is-3-colorable} routine. This routine admits a special parameter $\alpha$ that controls the level of recursive calls. In order to analyze its complexity, the recursion is fixed to $\alpha = 0$, and once the complexity for $\alpha = 0$ is obtained, the complexity for $\alpha > 0$ is established.

The \texttt{is-3-colorable} routine depends on the complexity of the contraction step (Step 1). At first sight, the contraction Step 1 has complexity of order $O(n^4)$ since it explores each $K_{112}$ subgraph whose number may increase with an increase in the number of combinations of four elements in the vertices of $G$. However, a relatively in-depth analysis reveals that the algorithm performs a vertex contraction until there is no other $K_{112}$. This means that this operation is bounded by the number of edges of the complement of $G$, which has a quadratic $O(n^2)$ order in the number of vertices. Steps 2 and 3 are absorbed into Step 1. Hence, for $\alpha = 0$,

\begin{align}\label{eq::cc:k112}
    \mathrm{g}(\text{is-3-colorable)} &= {n\choose 2},\\
     &= O(n^2), \; \text{for} \; \alpha = 0.
\end{align}

For $\alpha > 0$, the complexity of \texttt{is-3-colorable} routine also depends on recursive calls inside a for loop through every non-edge that has order $O(n^2)$; therefore, for $\alpha = 1$, we will have $O(n^2)O(n^2)$.

\begin{align}\label{eq::cc:k112b}
    \mathrm{g}(\text{is-3-colorable)} &= {n\choose 2}^{k+1},\\
     &= O(n^2)^{k + 1}.
\end{align}

Thus, on the basis of lemma~\ref{lemma:greedy:coloring}, it can be shown that the complexity of an algorithm that finds a 3-coloring is just one order higher.

\begin{equation}
\mathrm{g}(\text{\texttt{general-3COL}}) = O(n^2)^{k + 2}
\end{equation}

\section*{Experimental Results}
\label{sec:3}
The problem of evaluating algorithms experimentally could be very tricky if tests are performed on ``artificial instances,'' which may be uncorrelated or isolated from any specific practical application as claimed by Johnson~\cite{Johnson02} who proposed a methodological approach to the experimental analysis of algorithms. Nevertheless, there are some lines of research suggesting special distributions of graph instances on which purported NP-complete problem solvers should be evaluated in order to appropriately determine their performances (e.g., \cite{selman1996,Culberson2001,mizuno2008constructive}).

In the experimental part of this article, the algorithm was thoroughly evaluated over significant samples pertaining to three different graph distributions. Each class, and each distribution, has a good justification:
\begin{enumerate}
  \item Pseudo-random planar graphs~\cite{denise1996,bodirsky2003}. Planarity imposes some interesting structural properties, i.e., the 3-coloring problem on planar graphs is the only unqualified problem that remains open~\cite{steinberg1993state} since 1-coloring is trivial, 2-coloring is well characterized, and the maximum chromatic number on the plane is four~\cite{appel1977I,appel1977II}, and at the same time, the determination of 3-colorability of planar graphs is NP-complete~\cite{Stockmeyer1973,GJ79}.
  \item Random 4-regular planar graphs~\cite{manca1979,lehel1981,broersma1993}. Even more, the 3-colorability of four-regular planar graphs still remains NP-complete~\cite{Dailey1980289}, and most importantly, in this class, the average degree is fixed, and hence, the phase-transition phenomenon as defined for random graphs cannot be applied directly in this case.
  \item Erd\H{o}s-R\H{e}nyi connected random graphs~\cite{erdos1960}. Finally, sampling from the Erd\H{o}s-R\H{e}nyi (connected) random graphs distribution gives the necessary theoretical support for evaluating an algorithm in the general case, validating the theoretical bounds and allowing one to obtain results that can be compared against other algorithms in the literature, e.g., the best-performing 3-coloring algorithms proposed in the literature~\cite{Malaguti2010}.
\end{enumerate}

\subsection*{Sample generation details}
Random planar graphs~\cite{denise1996,bodirsky2003} are complex to generate, and their definitions and sampling methods are more difficult to implement. Instead, we opted for a relatively simple approach of generating ``pseudo random planar graphs.'' The procedure involves the generation of a maximal planar graph and the uniform selection of edges from this graph at random (i.e., with equal probability) to create another graph called a pseudo random graph. For this purpose, we used the ``Create Random Planar Graph'' algorithm implementation used by \cite{hochstattler2010catbox} (Gato - the Graph Animation Toolbox\footnote{CATBox: \url{http://schliep.org/CATBox}, Gato: \url{http://gato.sf.net}.}) for the creation of such random planar graphs. Only one modification was included to avoid generating a considerably large number of graphs containing a $K_4$ subgraph. The idea involves the generation of a $K_4$-free planar graph during 100 attempts returning the first encountered $K_4$-free graph; otherwise, returning the 100$^{th}$-generated graph.

Random 4-regular planar graphs are also very complex to generate. Here, the procedures described in Refs.~\cite{manca1979}, \cite{lehel1981} and \cite{broersma1993} to generate all the 4-regular planar graphs have been used. In particular, Theorem 2 of \cite{broersma1993} is used for generating 4-regular planar graphs. However, there is no theory defining a random 4-regular planar graph, so an ad hoc distribution has been specified to balance the proportion of Yes/No instances. The distribution has been obtained by assigning a probability to each graph transformation (see \cite{broersma1993}): $P(\overline{\phi}_A) = .80$, $P(\overline{\phi}_B) = .05$, $P(\overline{\phi}_C) = .10$, and $P(\overline{\phi}_F) = .05$.

The Erd\H{o}s-R\H{e}nyi~\cite{erdos1960} random graph, is a very well-known model that allows uniform sampling from graphs at random by specifying either a number of vertices and edges or the probability of a number of vertices and the edges. We followed standard methods to sample from this distribution. The only modification is that the generation of a connected graph is assured by first generating a simple path passing through all the vertices and then adding the remaining random edges.

The experiments are designed to study the behavior (not just the performance or running time) of the proposed algorithm. For this purpose, there are curves evaluating the algorithm's performance over a particular graph distribution and there is an initial comparative plot against the backtracking algorithm. The use of backtracking is restricted to the study of the algorithm's scalability since it is not possible to use backtracking consistently beyond the 100 vertex barrier because of its exponential growth. For planar graphs, the planar versions were used, while for random graphs, the improved versions were used.

All experiments were developed in the Python\footnote{\url{http://www.python.org/}} programming language using its standard libraries and the other libraries developed by the current author. Other software includes the planarity library\footnote{\url{http://code.google.com/p/planarity}}~\cite{boyer2004cutting} as well as the above mentioned Gato (the Graph Animation Toolbox) libraries. The experiments were realized on common personal computers, and no parallelism was used. All time measurements are done in seconds using Python's \texttt{time.clock()}\footnote{See \url{http://docs.python.org/library/time.html}.} function.

\subsection*{Experiment 1: scaling factor compared against backtracking}

The first part of the experimental analysis is a comparison between the scaling factor of the proposed algorithm with that of simple backtracking, in order to determine the differences in the behavior of both algorithms. This experiment involved the generation of uniformly random planar graph instances from 10 to 100 vertices (incremented by 1 and generating 100 graphs for each number) and solving each instance with both algorithms. Table~\ref{table:scaling} shows the parameters of the sample used in the experiment.

\begin{figure}[tb]
\begin{center}
\includegraphics[width=6in]{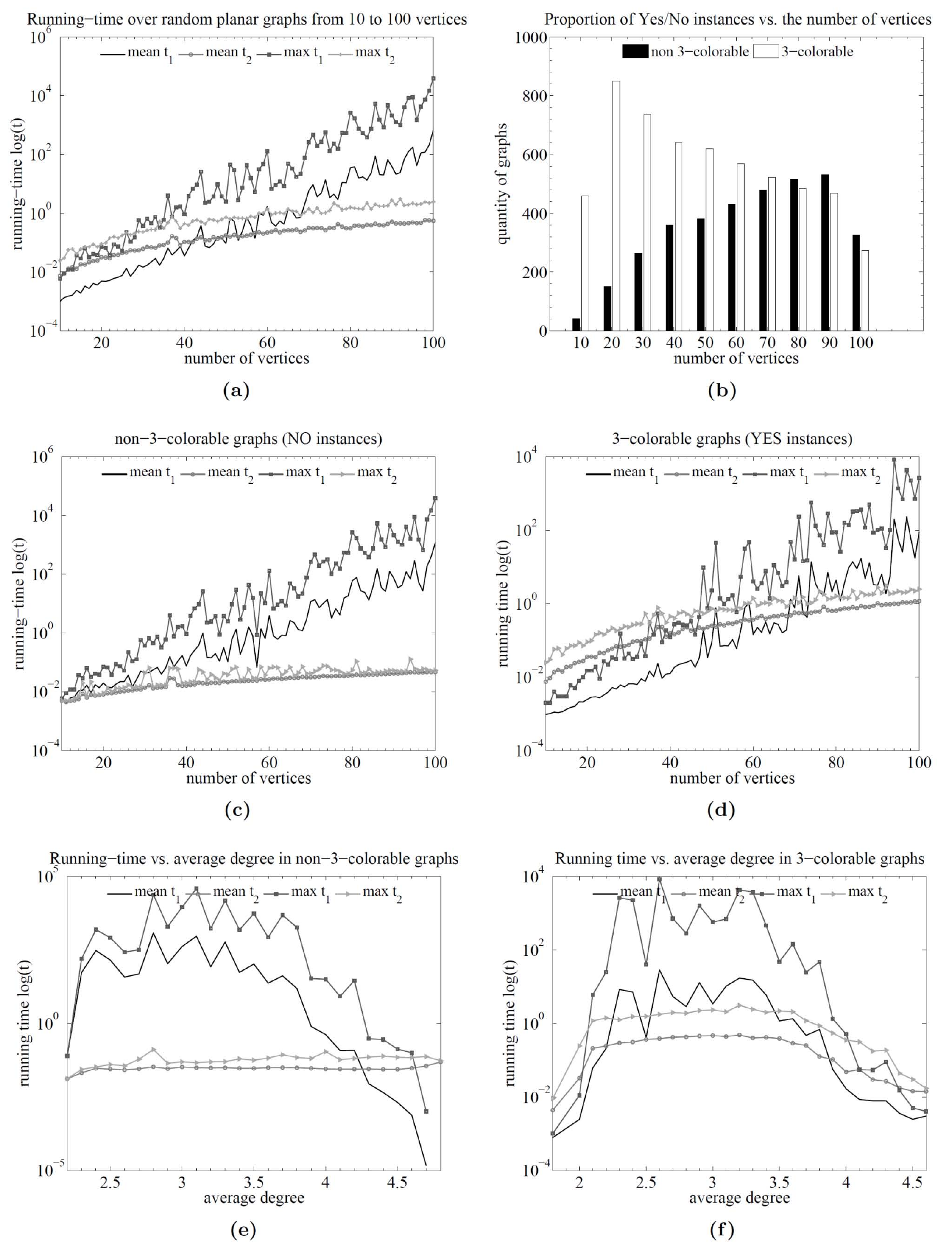}
\end{center}
\caption{ {\bf Backtracking vs. proposed algorithm.} Runtime analysis of the backtracking (brute force plus heuristic) ($t_1$) vs. the proposed parametric algorithm ($t_2$) over random planar graphs between 10 and 100 vertices. Plot (a) shows the running times as a function of the number of vertices for both kinds of instance types and for both algorithms. Plot (b) shows the proportion of 3-colorable and non-3-colorable graphs over the total number of graphs per number of vertices. Plots (c) and (d) also show the running times as a function of the number of vertices discriminated by the instance types (yes-or-no) so that subtle differences can be observed. Plots (e) and (f) also show the running times but as a function of the average degree and instance type.}
\label{graph:scaling}
\end{figure}

For each algorithm, the mean and maximum (max) running times were recorded as well as some other relevant statistics. Times labeled as $t_1$ correspond to the backtracking algorithm, while the $t_2$ times correspond to the proposed parametric algorithm. Comparative plots are shown in Figure~\ref{graph:scaling}; there are six plots from (a) to (f).

Figure~\ref{graph:scaling}a shows the running times as a function of the number of vertices for both kinds of instance types and for both algorithms. Figure~\ref{graph:scaling}b shows the proportion of 3-colorable and non-3-colorable graphs over the total number of graphs per number of vertices. It can be seen that the distribution tends to be uniform. Figures~\ref{graph:scaling}c and \ref{graph:scaling}d also show the running times as a function of the number of vertices but discriminated by the instance types (Yes/No) so that subtle differences can be observed.

The general results indicate that there is a crossing point in which backtracking continues to grow exponentially while the proposed algorithm remains polynomial (cf. Figure~\ref{graph:scaling}a at around 50 vertices); hence, a clear difference in the behavior of both algorithms is observed. This difference is clearer in the non-3-colorable instances (cf. Figure~\ref{graph:scaling}c) where the maximum running times of the parametric algorithm are relatively low in all the cases. Nevertheless, for the 3-colorable instances (cf. Figure~\ref{graph:scaling}d), the difference starts to be clear around graphs on 50 vertices.

Moreover, when running times are compared as a function of the average degree, there is a significant difference in the behavior of both algorithms. For non-3-colorable instances, the parametric algorithm exhibits an almost constant performance (cf. Figure~\ref{graph:scaling}e) and a totally uncorrelated curve against backtracking, which on the contrary is very sensitive to the average degree. This difference, although to a slightly minor degree, can also be observed in the 3-colorable instances as shown in Figure~\ref{graph:scaling}f.

\clearpage 

\subsection*{Experiment 2: random planar graphs}

This experiment involves the generation of uniformly random planar graph instances from 100 to 1000 vertices (incremented by 100 and generating 1000 graphs for each number) and solving each instance with the parametric algorithm. Table~\ref{table:randomplanar} shows the parameters of the sample used in the experiment. It is not possible to compare the results against backtracking because of its exponentially increasing running time.

\begin{figure}[tb]
\begin{center}
\includegraphics[width=6in]{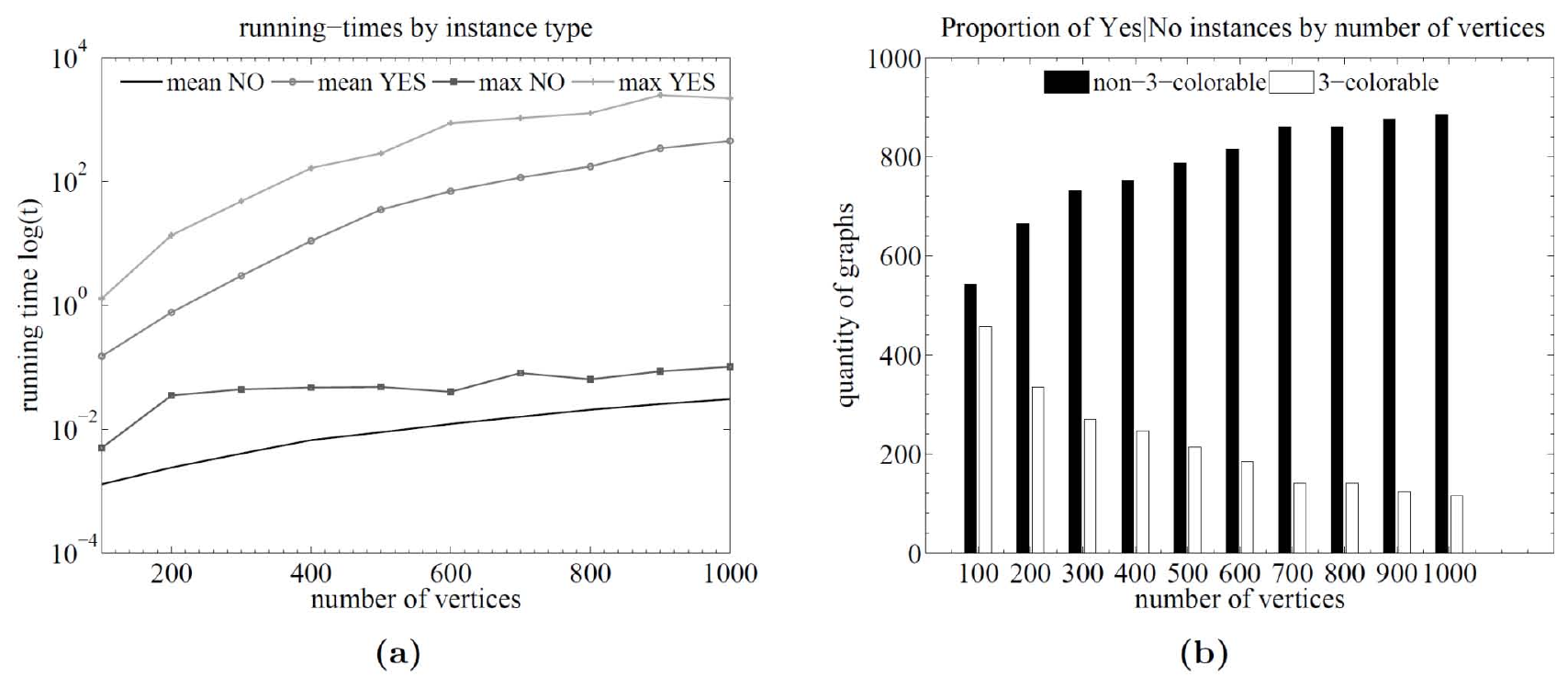}
\end{center}
\caption{{\bf Results for planar graphs.} Runtime analysis over random planar graphs between 100 and 1000 vertices (incremented by 100 and generating 1000 graphs for each number). Plot (a) shows the running times as a function of the number of vertices for both kinds of instance types. Plot (b) shows the proportion of 3-colorable and non-3-colorable graphs over the total number of graphs per number of vertices.}
\label{graph:random:planar}
\end{figure}

\begin{figure}[tb]
\begin{center}
\includegraphics[width=6in]{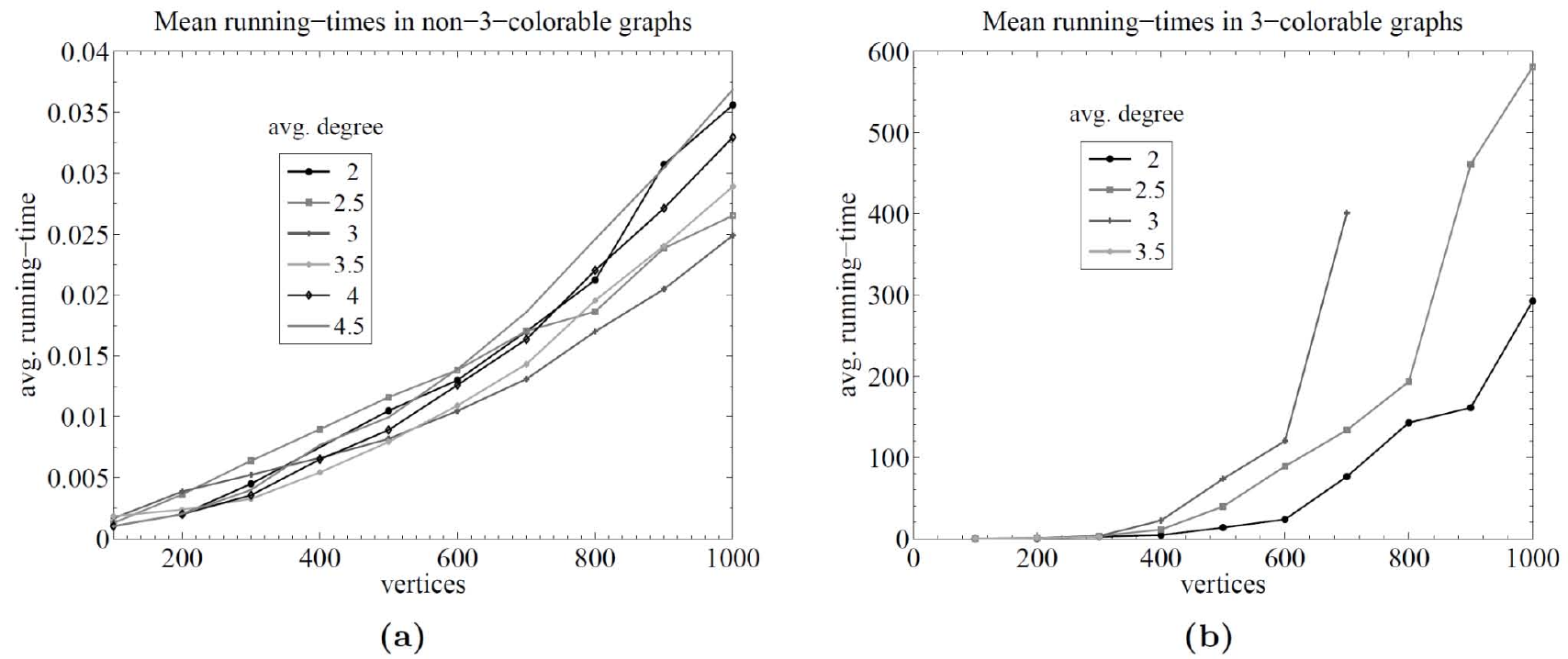}
\end{center}
\caption{{\bf Results for planar graphs.} Runtime analysis over random planar graphs considering instance type and average degree. Plots (a) and (b) also show the running times as a function of the number of vertices but discriminated by the average degree.}
\label{graph:random:planar2}
\end{figure}

For each instance type (Yes/No), the mean and maximum (max) running times where recorded, as well as some other relevant statistics. Comparative plots are shown in Figures \ref{graph:random:planar} and \ref{graph:random:planar2}.

Figure~\ref{graph:random:planar}a shows the running times as a function of the number of vertices for both kinds of instance types. Figure~\ref{graph:random:planar}b shows the proportion of 3-colorable and non-3-colorable graphs over the total number of graphs per number of vertices. It can be seen that the distribution is far from uniform. Figures~\ref{graph:random:planar2}a and \ref{graph:random:planar2}b also show the running times as a function of the number of vertices but discriminated by the average degree.

The results indicate that there is a difference in running times depending on the instance type (cf. Figure~\ref{graph:random:planar}a). This difference is expected since the proposed algorithm returns earlier (without entering the main loop) when a 3-uncolorability certificate is found. Even in the case when the average degree is considered, the difference is high (cf. Figures \ref{graph:random:planar2}a and \ref{graph:random:planar2}b).

\newpage

\subsection*{Experiment 3: random planar 4-regular graphs}

In this experiment, graphs were sampled from an ad-hoc distribution over the 4-regular planar graphs. The samples were used for generating graph instances from 100 to 1000 vertices (incremented by 100 and generating 1000 graphs for each number) and solving each instance with the parametric algorithm. Table~\ref{table:randomfour} shows the parameters of the samples used in the experiment.

\begin{figure}[tb]
\begin{center}
\includegraphics[width=6in]{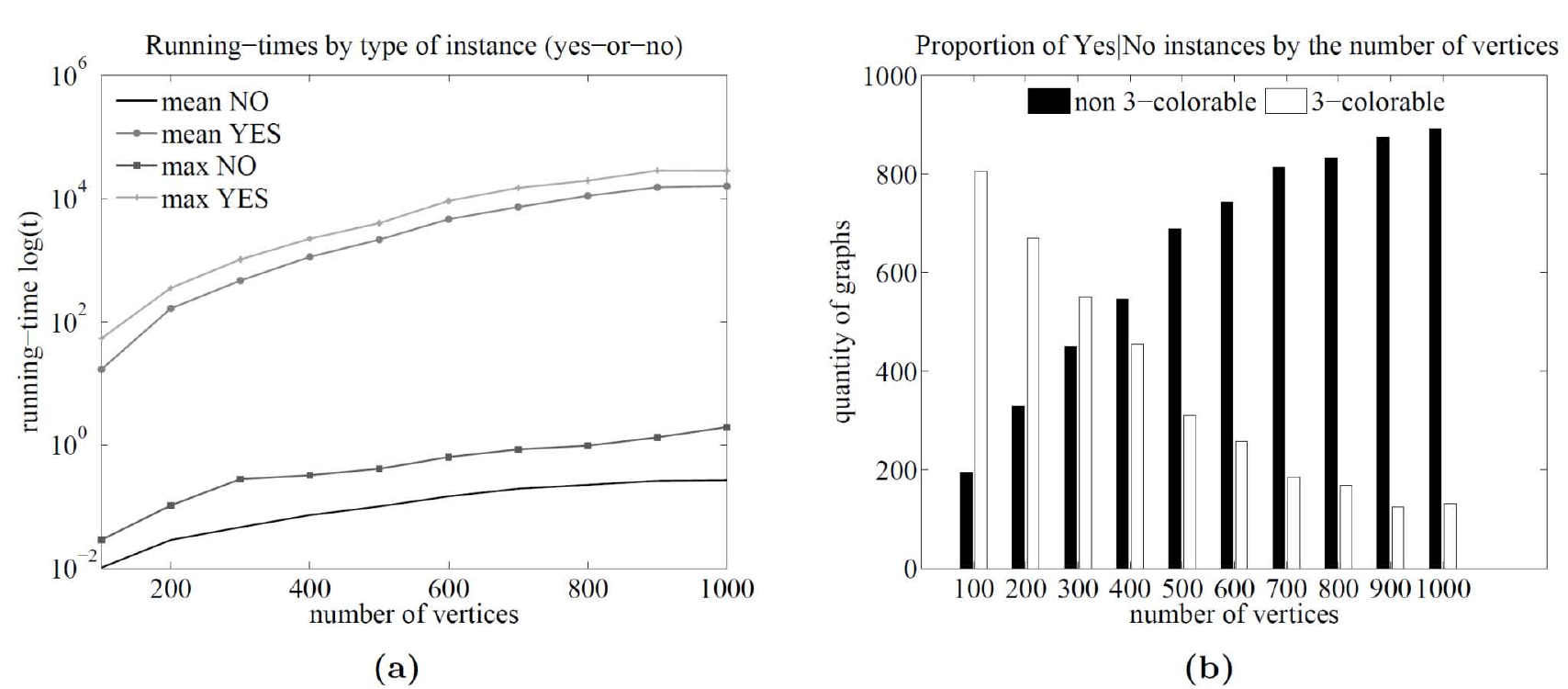}
\end{center}
\caption{{\bf Results for 4-regular planar graphs.} Runtime analysis over random 4-regular planar graphs between 100 and 1000 vertices. Plot (a) shows the running times as a function of the number of vertices for both kinds of instance types. Plot (b) shows the proportion of 3-colorable and non-3-colorable graphs over the total number of graphs per number of vertices.}
\label{graph:planar:regular}
\end{figure}

For each instance type (Yes/No), the mean and maximum (max) running times were recorded, as well as some other relevant statistics. Comparative plots are shown in Figure~\ref{graph:planar:regular}. Figure~\ref{graph:planar:regular}a shows the running times as a function of the number of vertices for both kinds of instance types. Figure~\ref{graph:planar:regular}b shows the proportion of 3-colorable and non-3-colorable graphs over the total number of graphs per number of vertices; it can be seen that the distribution is far from uniform.

These results indicate that there is a very significant difference in running times depending on the instance type (cf. Figure~\ref{graph:planar:regular}a). This difference was expected since the proposed algorithm returns earlier when a 3-uncolorability certificate is found. Again, the observed difference is high.

\clearpage

\subsection*{Experiment 4: Erd\H{o}s-R\H{e}nyi random graphs}

\begin{figure}[tb]
\begin{center}
\includegraphics[width=6in]{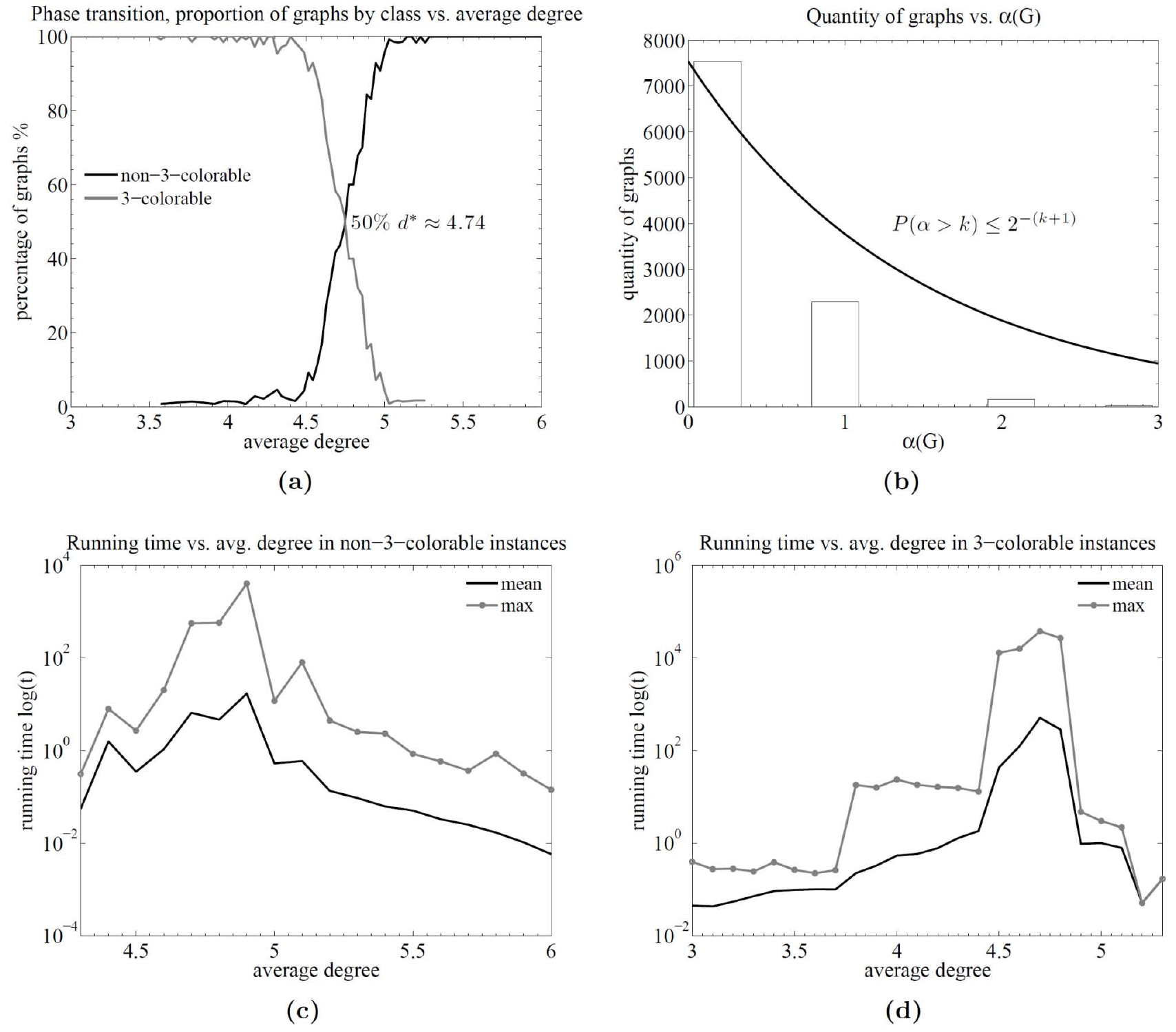}
\end{center}
\caption{{\bf Runtime analysis of the algorithm for random graphs}. The behavior of the proposed algorithm over the well-known Erd\H{o}s-R\H{e}nyi random graphs distribution. Plot (a) shows the quantity of 3-colorable and non-3-colorable graphs as a function of the average degree, i.e., a phase transition plot, in this case, occurring at around $d = 4.74$. Plot (b) shows the quantity of graphs corresponding to each $\alpha(G)$ value. As predicted by the theory, the proportion of graphs decreases exponentially as a function of $\alpha$ below the line of $2^{-(\alpha+1)}$. Plots (c) and (d) show the running times as a function of the average degree.}
\label{graph:randomgraps}
\end{figure}

In the last experiment, graphs were sampled from the well-known Erd\H{o}s-R\H{e}nyi random graphs distribution. The samples were graph instances of 100 vertices, generating in total 10000 graphs, and solving each instance with the parametric algorithm. Table~\ref{table:randomgraphs} shows the parameters of the sample used in the experiment. For each instance type (Yes/No), the mean and maximum (max) running times were recorded, as well as some other relevant statistics. Comparative plots are shown in Figure~\ref{graph:randomgraps}.

Figure~\ref{graph:randomgraps}a shows the quantity of 3-colorable and non-3-colorable graphs as a function of the average degree, i.e., a phase transition plot, in this case, occurring at around $d = 4.74$. It should be noted that this phase transition is for the \emph{connected} random graphs and not standard random graphs, which can contain many components, thus affecting the phase transition threshold. Figure~\ref{graph:randomgraps}b shows the quantity of graphs corresponding to each $\alpha(G)$ value. As predicted by the theory, almost all graphs have $\alpha(G) \leq k$ for some integer $k$, and the proportion of graphs decreases exponentially as a function of $\alpha$ clearly below the line of $2^{-(\alpha + 1)}$. These results confirm the established theoretical bounds.

Figures~\ref{graph:randomgraps}c and \ref{graph:randomgraps}d show the running time as a function of the average degree. It can be observed that in the random graphs case, the difference in running time is not as high as the difference observed in the planar graphs case. Further, there is a difference in the location of the harder instances for each kind of instance type: the harder instances for the non-3-colorable case are located around an average degree of $d\approx 5$ while, in the 3-coloring case, they are located slightly below an average degree of $d\approx 4.8$. Although the numbers seems to be very close, the shapes of the running-time curves are not. The shape of the running-time curve in Figure~\ref{graph:randomgraps}d falls sharply after 5, while the shape of the running-time curve in \ref{graph:randomgraps}c does not. This may indicate that there is a true difference in the location of the harder instances depending on the type (Yes/No) of the instance, and (to the best of my knowledge), there are no other works identifying a separation of a complexity threshold on the basis of the type of instance.

As observed in the experiments, the value of $\alpha(G)$ was directly correlated with the average degree $d = 2m/n$ (\emph{m} = edges, \emph{n} = vertices); hence, near the phase transition threshold ($d^*$)~\cite{mulet2002,Boettcher2004}, the probability of a relatively high value of $\alpha(G)$ increases.
\begin{equation}
d^* \approx
\left\{
  \begin{array}{ll}
    4.69,    & \hbox{\cite{mulet2002};} \\
    4.703,   & \hbox{\cite{Boettcher2004}}
  \end{array}
\right.
\end{equation}

However, many interesting questions remain open; e.g.,
\begin{itemize}
  \item What is the exact distribution of $\alpha(G)$ in random graphs?
  \item Apart from the average degree, what other parameters are related to $\alpha(G)$?
  \item Given an arbitrary input graph $G$, can the $\alpha(G)$ value be predicted (exactly or approximately) and what is the best possible approximation to $\alpha(G)$?
  \item As for the chromatic number $\chi(G)$, is there any (efficient) graph construction mechanism that allows the generation of graphs with arbitrarily large $\alpha(G)$?
\end{itemize}

\clearpage

\section*{Discussion}
\label{sec:conclusion}

In this article, an asymptotic parametric exact 3-coloring algorithm has been presented. This is (to the best of my knowledge) the first algorithm of its kind for the 3-coloring problem.

The maximal complexity of the algorithm is controlled by the parameter ($\alpha$) that bounds the recursion depth and determines its running time. The algorithm relies on the efficient search of 3-uncolorability certificates. Here, a formal definition of the 3-uncolorability certificate has been introduced. This is the central theoretical concept that allowed the development of the proposed algorithm. The definition of the 3-uncolorability certificate presented here is (to the best of my knowledge) the first one that is formally presented and the most naturally related to the 3-coloring problem.

A very significant feature of 3-uncolorability certificates is that it is possible to obtain them from small subgraphs of a particular graph, indeed, as small as four vertices (i.e. by finding a $K_4$ subgraph). Hence, an interesting theoretical analysis that should follow is to study of the behavior of $\alpha(G)$ on 4-critical graphs since in this class, there is no subgraph with chromatic number four, and hence, finding unavoidable vertex contractions may be relatively hard (e.g., see Ref.~\cite{mizuno2008constructive} for a good initial development of this idea). Hence, a classification of 4-critical graphs on the basis of $\alpha(G)$ can lead to very significant results.

There is an interesting symmetry between coloring and uncolorability certificates:

\begin{itemize}
  \item In order to show that a graph is 3-colorable, it is sufficient to encounter just one legal coloring; nevertheless, any legal coloring must assign a color to all the vertices of the graph without violating any constraint since it remains hard to determine if a partial coloring is extensible to all the vertices of the graph.
  \item Instead, in order to show that a graph is not 3-colorable, one needs to verify that none of the possible 3-colorings is a legal one; nevertheless, for obtaining a 3-uncolorability certificate, it is sufficient to encounter just one non-3-colorable subgraph (e.g., a 4-critical subgraph), i.e., a small graph.
\end{itemize}

Thus, while for considerably large graphs, just verifying a legal 3-coloring can be complex in practice, it remains practical (at least in theory) to determine 3-uncolorability even for such graphs.

Hence, in principle, finding uncolorability certificates can be assumed to be at least of the same kind as finding colorings. Thus, there should not be any problem in the development of 3-coloring algorithms on the basis of a search for 3-uncolorability certificates that eventually reach the same level of sophistication and performance as its coloring-based peers.

Moreover, if the algorithm is used as a heuristic, e.g., to test whether a solution can be found quickly (``just by chance'') with a relatively low (efficient) $\alpha$, the algorithm will search for both 3-colorings and 3-uncolorability certificates at the same time, in clear contrast with the use of backtracking, greedy-based, and randomized 3-coloring algorithms. Further, this feature is particularly important as its consideration ensure that it is not necessary to trust the correctness of the algorithm itself or the particular implementation used in order to recognize that the provided solution is correct since the result can be efficiently verified using just the solution provided, i.e., a legal 3-coloring or a 3-uncolorability certificate.

The developed theoretical analysis guarantees some good features of the proposed algorithm. The most important one, for both practical and theoretical purposes, is that while the algorithm relies on the value of $\alpha$ to be able to find a certificate, the probability that $\alpha(G) > k$ decreases at the rate $P(\alpha(G) > k) \leq 2^{-(k + 1)}$, e.g., for $k = 19$, there is less chance than one in a million of not obtaining a solution with the proposed polynomial algorithm (i.e., probability of success = 0.999999), assuming that the input is a random graph.

Thus, while certainly beyond some value of $\alpha$, the running times would become prohibitive given the current state of the computing machinery, the developed algorithm scales polynomially, and the probability of obtaining a solution (success) grows exponentially with an increment of $\alpha$. Hence, any step (i.e., any investment) in computing power technology will lead to a huge (exponential) growth of the class of tractable 3-coloring instances, as well as CSPs in general.

Perhaps, it could be the case that we can achieve at least a ``technological tractability''?, i.e., a guaranteed number of instances such that almost all computational problems of practical interest could be solved for $\alpha(G) \in [0,k]$ for some integer $k$.

It should also be observed that increasing $\alpha$ as the result of technological progress implies that $\alpha \neq f(\text{input})$, i.e., $\alpha$ is not a function of the input. Does technological progress imply a polynomial algorithm for 3-colorability?

In addition, since 3-colorability is NP-complete and to each graph corresponds a unique $\alpha(G)$, a classification based on $\alpha(G)$ of all the NP-complete problem instances can be done by a reduction of each problem instance to a 3-coloring instance $G$ such that $\alpha(G)= k$ for some $k\in \mathbb{N}$.

However, can we define NP as follows? Let us define $\text{NP}(\alpha)$ as the class of problems in NP that are also in P for some particular value of $\alpha(G) \in \mathbb{N}$. Then,

\begin{align}
\text{NP} & = \bigcup_{\alpha=0}^{\infty} \text{NP}(\alpha),
\end{align}
i.e., can then NP be defined as the infinite union of problems in P?

Finally, even determining the infiniteness of $\alpha(G)$, is there, as in the case of the maximal degree four, $(\triangle(G) \leq 4)$; a $k\in\mathbb{N}$ such that determining 3-colorability over a class of graphs with $\alpha(G) \leq k$ is still NP-complete, i.e., P = NP?

In the maximal-degree case, we know that 3-colorability restricted to $\triangle(G) \leq 4$ is still NP-complete. Nevertheless, the problem is to determine whether a polynomial algorithm exists or not.

On the contrary, in the finite-$\alpha(G)$ case, we know that 3-colorability restricted to $\alpha(G) \leq k$ is in P. Nevertheless, the problem is to determine whether it is NP-complete for a class of graphs and finite $k\in\mathbb{N}$.

\section*{Reproducibility note}

The working source-code of the algorithm and all the software libraries needed to appropriately use and experiment with the algorithm have been released and are available at the publisher's website.

Furthermore, there is a web application that implements the algorithm inside the Google App Engine cloud computing framework. The users can visit the site and test the algorithm at the following url:
\begin{itemize}
  \item \url{http://graph-coloring.appspot.com}
\end{itemize}

The web coloring application just asks for a file where a graph is defined following the plain text version of the simple edge-list according to the DIMACS standard format specification (\url{http://mat.gsia.cmu.edu/COLOR/general/ccformat.ps}), such as the .col files in \url{http://mat.gsia.cmu.edu/COLOR/instances.html}.

%% file: tables.tex
\begin{table}[!ht]
\caption{\bf{A priori expected performance of the algorithm with respect to 3-colorability and graph density parameters}. \label{table:analysis1}}
\begin{tabular}{p{2cm} p{6cm} p{6cm}}
  \toprule
  \textbf{KIND} & \textbf{3-COLORABLE} & \textbf{NOT 3-COLORABLE} \\ \toprule %
  & & \\
  \textbf{SPARSE} $(d<d^*)$ & High probability of a short running time due to the existence of many legal colorings. & High probability of a short running time since almost all non-3-colorable graphs contain a $K_4$ and hence the probability of obtaining a $K_4$-free non-3-colorable graph decreases rapidly when the average degree falls below the phase transition threshold. \\
 & & \\ 
$d^*$ & Harder instances. & Harder instances. \\
 & & \\ 
  \textbf{DENSE} $(d>d^*)$ & High probability of a short running time due to the existence of many $K_{112}$ subgraphs that prune the search, e.g., graphs tend to be uniquely colorable. & High probability of a short running time due to the existence of many small 3-uncolorability certificates due to the average degree, e.g., too many $K_4$-subgraphs.\\
& & \\
  \bottomrule
\end{tabular}
\begin{flushleft} Average case (expected) performance of the algorithm with respect to the density of the graph, i.e., above/below the phase transition threshold ($d^* \simeq 4.69$) and the type of instance (Yes/No).
\end{flushleft}
\end{table}

\begin{table}[!ht]
\caption{{\bf Parameters of the scalability test between backtracking and the proposed algorithm.} \label{table:scaling}}
\begin{tabular}{p{4cm} p{10.0cm}}
  \toprule
  \textbf{Sample property} & \textbf{Value} \\  \toprule
  Sample type: & Random planar graphs. \\ \midrule
  Sample size: & 9000 graphs. \\ \midrule
  Vertex number: & From 10 to 100 vertices, incremented by 1. \\ \midrule
  Average degree: & From 2 to 5 edges uniformly distributed. \\ \midrule
  Group size: & 100 graphs per number of vertices. \\    \bottomrule
\end{tabular}
\begin{flushleft} Random planar graph instances from 10 to 100 vertices incremented by 1 and generating 100 graphs for each number of vertices, i.e., 9000 graphs in total.
\end{flushleft}
\end{table}

\begin{table}[!ht]
\caption{{\bf Parameters of the sample used in the random planar graphs test}. \label{table:randomplanar}}
\begin{tabular}{p{4cm} p{10cm}}
  \toprule
  \textbf{Sample property} & \textbf{Value} \\  \toprule
  Sample type: & Random planar graphs. \\ \midrule
  Sample size: & 10000 graphs. \\ \midrule
  Vertex number: & From 100 to 1000 vertices, incremented by 100. \\ \midrule
  Average degree: & From 2 to 5 edges uniformly distributed. \\ \midrule
  Group size: & 100 graphs per number of vertices. \\    \bottomrule
\end{tabular}
\begin{flushleft} Uniformly random planar graph instances from 100 to 1000 vertices incremented by 100 and generating 1000 graphs for each number of vertices, i.e., 10000 graphs in total.
\end{flushleft}
\end{table}

\begin{table}[!ht]
\caption{{\bf Parameters of the sample used in the random planar 4-regular graphs test.} \label{table:randomfour}}
\begin{tabular}{p{4cm} p{10cm}}
  \toprule
  \textbf{Sample property} & \textbf{Value} \\  \toprule
  Sample type: & Random planar 4-regular graphs. \\ \midrule
  Sample distribution: & $P(\overline{\phi}_A) = .80$, $P(\overline{\phi}_B) = .05$, $P(\overline{\phi}_C) = .10$ and $P(\overline{\phi}_F) = .05$ \\ \midrule
  Sample size: & 10000 graphs. \\ \midrule
  Vertex number: & From 100 to 1000 vertices, incremented by 100. \\ \midrule
  Degree: & Fixed: 4-regular graphs. \\ \midrule
  Group size: & 1000 graphs per number of vertices. \\    \bottomrule
\end{tabular}
\begin{flushleft} Graphs are sampled from an ad-hoc distribution over the 4-regular planar graphs. The sample consist of graph instances from 100 to 1000 vertices incremented by 100 and generating 1000 graphs for each number of vertices, i.e., 10000 graphs in total. For the exact meaning of each graph transformation operation, i.e., $\overline{\phi}_A, \overline{\phi}_B, \overline{\phi}_C  \; and \;  \overline{\phi}_F$ see~Ref.~\cite{broersma1993}.
\end{flushleft}

\end{table}

\begin{table}[!ht]
\caption{{\bf Parameters of the sample used in the random graphs test.} \label{table:randomgraphs}}
\begin{tabular}{p{4cm} p{10cm}}
  \toprule
  \textbf{Sample property} & \textbf{Value} \\  \toprule
  Sample type: & Erd\H{o}s-R\H{e}nyi random graphs. \\ \midrule
  Sample size: & 10000 graphs. \\ \midrule
  Vertex number: & Fixed: 100 vertices. \\ \midrule
  Average degree: & From 3 to 6 edges uniformly distributed. \\ \bottomrule
\end{tabular}
\begin{flushleft} Graphs are sampled from the well-known Erd\H{o}s-R\H{e}nyi random graphs distribution. The sample consist of graph instances for 100 vertices, generating in total 10000 graphs. For each number of vertices, the average degree is varied from 3 to 6.
\end{flushleft}
\end{table}

%% file: ParametricColoringarXiv.bbl
\begin{thebibliography}{10}
\expandafter\ifx\csname url\endcsname\relax
  \def\url#1{\texttt{#1}}\fi
\expandafter\ifx\csname urlprefix\endcsname\relax\def\urlprefix{URL }\fi
\expandafter\ifx\csname href\endcsname\relax
  \def\href#1#2{#2} \def\path#1{#1}\fi

\bibitem{jones2008artificial}
M.~Jones, \href{http://books.google.es/books?id=ekUHwvRP7nUC}{{Artificial
  Intelligence: A Systems Approach}}, Computer Science, Jones \& Bartlett
  Publishers, Incorporated, 2008.
\newline\urlprefix\url{http://books.google.es/books?id=ekUHwvRP7nUC}

\bibitem{Wigderson1983}
A.~Wigderson, \href{http://doi.acm.org/10.1145/2157.2158}{Improving the
  performance guarantee for approximate graph coloring}, Journal of the ACM
  (JACM) 30 (1983) 729--735.
\newblock \href {http://dx.doi.org/http://doi.acm.org/10.1145/2157.2158}
  {\path{doi:http://doi.acm.org/10.1145/2157.2158}}.
\newline\urlprefix\url{http://doi.acm.org/10.1145/2157.2158}

\bibitem{park1996}
T.~Park, C.~Lee, Application of the graph coloring algorithm to the frequency
  assignment problem, Journal of the Operations Research Society of Japan-Keiei
  Kagaku 39~(2) (1996) 258--265.

\bibitem{ramani2004}
A.~Ramani, F.~Aloul, I.~Markov, K.~Sakallah, Breaking instance-independent
  symmetries in exact graph coloring, in: Proceedings of the conference on
  Design, automation and test in Europe-Volume 1, IEEE Computer Society, 2004,
  p. 10324.

\bibitem{zdeborova2007}
L.~Zdeborov\'a, F.~Krzakala, Phase transitions in the coloring of random
  graphs, Physical Review E 76~(3) (2007) 031131.

\bibitem{Ingrid2005}
I.~Abfalter,
  \href{http://www.tbi.univie.ac.at/papers/Abstracts/ingrid_diss.pdf}{Nucleic
  acid sequence design as a graph colouring problem}, Ph.D. thesis,
  Universit\"{a}t Wien (November 2005).
\newline\urlprefix\url{http://www.tbi.univie.ac.at/papers/Abstracts/ingrid_diss.pdf}

\bibitem{pevzner1995open}
P.~Pevzner, M.~Waterman, Open combinatorial problems in computational molecular
  biology, in: Theory of Computing and Systems, 1995. Proceedings., Third
  Israel Symposium on the, IEEE, 1995, pp. 158--173.

\bibitem{karp2011heuristic}
R.~Karp, Heuristic algorithms in computational molecular biology, Journal of
  Computer and System Sciences 77~(1) (2011) 122--128.

\bibitem{Arora2009}
S.~Arora, B.~Barak, Computational Complexity: A Modern Approach, 1st Edition,
  Cambridge University Press, New York, NY, USA, 2009.

\bibitem{goldreich2008}
O.~Goldreich, \href{http://books.google.es/books?id=EuguvA-w5OEC}{Computational
  complexity: a conceptual perspective}, Cambridge University Press, 2008.
\newline\urlprefix\url{http://books.google.es/books?id=EuguvA-w5OEC}

\bibitem{cook1971}
S.~A. Cook, \href{http://doi.acm.org/10.1145/800157.805047}{The complexity of
  theorem-proving procedures}, in: Proceedings of the third annual ACM
  symposium on Theory of computing, STOC '71, ACM, New York, NY, USA, 1971, pp.
  151--158.
\newblock \href {http://dx.doi.org/http://doi.acm.org/10.1145/800157.805047}
  {\path{doi:http://doi.acm.org/10.1145/800157.805047}}.
\newline\urlprefix\url{http://doi.acm.org/10.1145/800157.805047}

\bibitem{karp1972reducibility}
R.~M. Karp, Reducibility among combinational problems, Complexity of Computer
  Computations (1972) 85--103.

\bibitem{levin1973universal}
L.~Levin, Universal sequential search problems, Problemy Peredachi Informatsii
  9~(3) (1973) 115--116.

\bibitem{GJ79}
M.~R. Garey, D.~S. Johnson, Computers and Intractability, A Guide to the Theory
  of NP-Completeness, W.H. Freeman and Co., San Francisco, 1979.

\bibitem{steinberg1993state}
R.~Steinberg, The state of the three color problem, Annals of discrete
  mathematics 55 (1993) 211--248.

\bibitem{hogg1996phase}
T.~Hogg, B.~Huberman, C.~Williams, Phase transitions and the search problem,
  Artificial intelligence 81~(1-2) (1996) 1--15.

\bibitem{Culberson2001}
J.~Culberson, I.~Gent,
  \href{http://portal.acm.org/citation.cfm?id=500494.500504}{Frozen development
  in graph coloring}, Theoretical computer science 265 (2001) 227--264.
\newblock \href {http://dx.doi.org/10.1016/S0304-3975(01)00164-5}
  {\path{doi:10.1016/S0304-3975(01)00164-5}}.
\newline\urlprefix\url{http://portal.acm.org/citation.cfm?id=500494.500504}

\bibitem{mulet2002}
R.~Mulet, A.~Pagnani, M.~Weigt, R.~Zecchina, Coloring random graphs, Physical
  review letters 89~(26) (2002) 268701.

\bibitem{Boettcher2004}
S.~Boettcher, A.~G. Percus, Extremal optimization at the phase transition of
  the three-coloring problem, Physical Review E 69~(6) (2004) 066703.
\newblock \href {http://dx.doi.org/10.1103/PhysRevE.69.066703}
  {\path{doi:10.1103/PhysRevE.69.066703}}.

\bibitem{erdos1960}
P.~Erd\H{o}s, A.~R\H{e}nyi, On the evolution of random graphs, Publications of
  tke Matkemafical Insfifufe of the Hungarian Academy of Sciences 5.

\bibitem{erdos1973asymptotic}
P.~Erdos, D.~Kleitman, B.~Rothschild, Asymptotic enumeration of kn-free graphs,
  in: Colloquio Internazionale sulle Teorie Combinatorie (Rome, 1973), Vol.~2,
  Atti dei Convegni Lincei, 17, Accad. Naz. Lincei, Roma, 1976, pp. 19--27.

\bibitem{Borodin1996}
O.~V. Borodin,
  \href{http://portal.acm.org/citation.cfm?id=228780.228793}{Structural
  properties of plane graphs without adjacent triangles and an application to
  3-colorings}, Journal of Graph Theory 21 (1996) 183--186.
\newblock \href
  {http://dx.doi.org/10.1002/(SICI)1097-0118(199602)21:2<183::AID-JGT7>3.0.CO;2-N}
  {\path{doi:10.1002/(SICI)1097-0118(199602)21:2<183::AID-JGT7>3.0.CO;2-N}}.
\newline\urlprefix\url{http://portal.acm.org/citation.cfm?id=228780.228793}

\bibitem{Borodin2005}
O.~V. Borodin, A.~Glebov, A.~Raspaud, M.~Salavatipour,
  \href{http://www.sciencedirect.com/science/article/pii/S0095895604001170}{Planar
  graphs without cycles of length from 4 to 7 are 3-colorable}, Journal of
  Combinatorial Theory, Series B 93~(2) (2005) 303 -- 311.
\newblock \href {http://dx.doi.org/DOI: 10.1016/j.jctb.2004.11.001}
  {\path{doi:DOI: 10.1016/j.jctb.2004.11.001}}.
\newline\urlprefix\url{http://www.sciencedirect.com/science/article/pii/S0095895604001170}

\bibitem{Wang2007}
W.-f. Wang, M.~Chen, \href{http://dx.doi.org/10.1007/s11425-007-0106-4}{Planar
  graphs without 4,6,8-cycles are 3-colorable}, Science in China Series A:
  Mathematics 50 (2007) 1552--1562, 10.1007/s11425-007-0106-4.
\newline\urlprefix\url{http://dx.doi.org/10.1007/s11425-007-0106-4}

\bibitem{Borodin2009}
O.~V. Borodin, A.~N. Glebov, M.~Montassier, A.~Raspaud,
  \href{http://portal.acm.org/citation.cfm?id=1537317.1537543}{Planar graphs
  without 5- and 7-cycles and without adjacent triangles are 3-colorable},
  Journal of Combinatorial Theory, Series B. 99 (2009) 668--673.
\newblock \href {http://dx.doi.org/10.1016/j.jctb.2008.11.001}
  {\path{doi:10.1016/j.jctb.2008.11.001}}.
\newline\urlprefix\url{http://portal.acm.org/citation.cfm?id=1537317.1537543}

\bibitem{Thomassen1994}
C.~Thomassen,
  \href{http://www.sciencedirect.com/science/article/pii/S0095895684710690}{Gr{\H{o}}tzsch's
  3-color theorem and its counterparts for the torus and the projective plane},
  Journal of Combinatorial Theory, Series B 62~(2) (1994) 268 -- 279.
\newblock \href {http://dx.doi.org/DOI: 10.1006/jctb.1994.1069} {\path{doi:DOI:
  10.1006/jctb.1994.1069}}.
\newline\urlprefix\url{http://www.sciencedirect.com/science/article/pii/S0095895684710690}

\bibitem{Johnson1974}
D.~S. Johnson,
  \href{http://www.sciencedirect.com/science/article/pii/S0022000074800449}{Approximation
  algorithms for combinatorial problems}, Journal of Computer and System
  Sciences 9~(3) (1974) 256 -- 278.
\newblock \href {http://dx.doi.org/DOI: 10.1016/S0022-0000(74)80044-9}
  {\path{doi:DOI: 10.1016/S0022-0000(74)80044-9}}.
\newline\urlprefix\url{http://www.sciencedirect.com/science/article/pii/S0022000074800449}

\bibitem{johnson1974worst}
D.~S. Johnson, Worst case behavior of graph coloring algorithms, in:
  Proceedings of the Fifth Southeastern Conference on Combinatorics, Graph
  Theory and Computing (Florida Atlantic Univ., Boca Raton, Fla., 1974, F.
  Hoffman et al., eds.), 1974, pp. 513--527.

\bibitem{Garey1976nearoptimal}
M.~R. Garey, D.~S. Johnson, \href{http://doi.acm.org/10.1145/321921.321926}{The
  complexity of near-optimal graph coloring}, Journal of the ACM (JACM) 23
  (1976) 43--49.
\newblock \href {http://dx.doi.org/http://doi.acm.org/10.1145/321921.321926}
  {\path{doi:http://doi.acm.org/10.1145/321921.321926}}.
\newline\urlprefix\url{http://doi.acm.org/10.1145/321921.321926}

\bibitem{berger1990better}
B.~Berger, J.~Rompel, A better performance guarantee for approximate graph
  coloring, Algorithmica 5~(1) (1990) 459--466.

\bibitem{halldorsson1993still}
M.~Halld{\'o}rsson, A still better performance guarantee for approximate graph
  coloring, Information Processing Letters 45~(1) (1993) 19--23.

\bibitem{blum1994new}
A.~Blum, New approximation algorithms for graph coloring, Journal of the ACM
  (JACM) 41~(3) (1994) 516.

\bibitem{Arora2006}
S.~Arora, E.~Chlamtac, \href{http://doi.acm.org/10.1145/1132516.1132548}{New
  approximation guarantee for chromatic number}, in: Proceedings of the
  thirty-eighth annual ACM symposium on Theory of computing, STOC '06, ACM, New
  York, NY, USA, 2006, pp. 215--224.
\newblock \href {http://dx.doi.org/http://doi.acm.org/10.1145/1132516.1132548}
  {\path{doi:http://doi.acm.org/10.1145/1132516.1132548}}.
\newline\urlprefix\url{http://doi.acm.org/10.1145/1132516.1132548}

\bibitem{karger1998approximate}
D.~Karger, R.~Motwani, M.~Sudan, Approximate graph coloring by semidefinite
  programming, Journal of the ACM (JACM) 45~(2) (1998) 246--265.

\bibitem{khanna2000hardness}
S.~Khanna, N.~Linial, S.~Safra, On the hardness of approximating the chromatic
  number, Combinatorica 20~(3) (2000) 393--415.

\bibitem{downey1999parameterized}
R.~Downey, M.~Fellows, Parameterized complexity, Vol.~5, Springer New York,
  1999.

\bibitem{flum2006parameterized}
J.~Flum, M.~Grohe, Parameterized complexity theory (texts in theoretical
  computer science. an eatcs series) (2006).

\bibitem{Boppana1987}
R.~B. Boppana, J.~Hastad, S.~Zachos, Does {co-NP} have short interactive
  proofs?, Information Processing Letters 25~(2) (1987) 127--132.
\newblock \href {http://dx.doi.org/10.1016/0020-0190(87)90232-8}
  {\path{doi:10.1016/0020-0190(87)90232-8}}.

\bibitem{Fortnow1988}
L.~Fortnow, M.~Sipser, Are there interactive proofs for {co-NP} languages?,
  Information Processing Letters 28 (1988) 249--251.

\bibitem{Bes2005}
J.-N. Bes, P.~Jegou, Proving graph un-colorability with a consistency check of
  {CSP}, in: Tools with Artificial Intelligence, 2005. ICTAI 05. 17th IEEE
  International Conference on, 2005, pp. 2 pp. --694.
\newblock \href {http://dx.doi.org/10.1109/ICTAI.2005.102}
  {\path{doi:10.1109/ICTAI.2005.102}}.

\bibitem{denise1996}
A.~Denise, M.~Vasconcellos, D.~Welsh, The random planar graph, Congressus
  numerantium (1996) 61--80.

\bibitem{bodirsky2003}
M.~Bodirsky, C.~Gr\H{o}pl, M.~Kang, Generating labeled planar graphs uniformly
  at random, Automata, Languages and Programming (2003) 191--191.

\bibitem{manca1979}
P.~Manca, Generating all planer graphs regular of degree four, Journal of graph
  theory 3~(4) (1979) 357--364.

\bibitem{lehel1981}
J.~Lehel, Generating all 4-regular planar graphs from the graph of the
  octahedron, Journal of graph theory 5~(4) (1981) 423--426.

\bibitem{broersma1993}
H.~Broersma, A.~Duijvestijn, F.~G\H{o}bel, Generating all 3-connected 4-regular
  planar graphs from the octahedron graph, Journal of graph theory 17~(5)
  (1993) 613--620.

\bibitem{JT95}
T.~R. Jensen, B.~Toft, Graph coloring problems, Wiley-Interscience Series in
  Discrete Mathematics and Optimization, John Wiley \& Sons, Chichester-New
  York-Brisbane-Toronto-Singapore, 1995.

\bibitem{Chartrand2008}
G.~Chartrand, P.~Zhang, Chromatic Graph Theory, 1st Edition, Chapman \&
  Hall/CRC, 2008.

\bibitem{Stockmeyer1973}
L.~Stockmeyer, Planar 3-colorability is polynomial complete, SIGACT News 5
  (1973) 19--25.
\newblock \href {http://dx.doi.org/10.1145/1008293.1008294}
  {\path{doi:10.1145/1008293.1008294}}.

\bibitem{martinh2011}
J.~A. Martin~H., Minimal non-extensible precolorings and implicit-relations,
  CoRR - Computing Research Repository abs/1104.0510 (2011) 1--10.

\bibitem{heawood1898four}
P.~Heawood, On the four-colour map theorem, Quarterly Journal of Pure and
  Applied Mathematics 29 (1898) 270--285.

\bibitem{Johnson02}
D.~S. Johnson, A theoretician's guide to the experimental analysis of
  algorithms, in: Data Structures, Near Neighbor Searches, and Methodology:
  Fifth and Sixth DIMACS Implementation Challenges, American Mathematical
  Society, 2002, pp. 215--250.

\bibitem{selman1996}
B.~Selman, D.~Mitchell, H.~Levesque, Generating hard satisfiability problems*
  1, Artificial intelligence 81~(1-2) (1996) 17--29.

\bibitem{mizuno2008constructive}
K.~Mizuno, S.~Nishihara, Constructive generation of very hard 3-colorability
  instances, Discrete Applied Mathematics 156~(2) (2008) 218--229.

\bibitem{appel1977I}
K.~Appel, W.~Haken, J.~Koch, Every planar map is four colorable. {Part I}:
  Discharging, Illinois Journal of Mathematics 21~(3) (1977) 429--490.

\bibitem{appel1977II}
K.~Appel, W.~Haken, J.~Koch, Every planar map is four colorable. {Part II}:
  Reducibility, Illinois Journal of Mathematics 21~(3) (1977) 491--567.

\bibitem{Dailey1980289}
D.~P. Dailey,
  \href{http://www.sciencedirect.com/science/article/pii/0012365X80902368}{Uniqueness
  of colorability and colorability of planar 4-regular graphs are np-complete},
  Discrete Mathematics 30~(3) (1980) 289 -- 293.
\newblock \href {http://dx.doi.org/DOI: 10.1016/0012-365X(80)90236-8}
  {\path{doi:DOI: 10.1016/0012-365X(80)90236-8}}.
\newline\urlprefix\url{http://www.sciencedirect.com/science/article/pii/0012365X80902368}

\bibitem{Malaguti2010}
E.~Malaguti, P.~Toth,
  \href{http://dx.doi.org/10.1111/j.1475-3995.2009.00696.x}{A survey on vertex
  coloring problems}, International Transactions in Operational Research 17~(1)
  (2010) 1--34.
\newblock \href {http://dx.doi.org/10.1111/j.1475-3995.2009.00696.x}
  {\path{doi:10.1111/j.1475-3995.2009.00696.x}}.
\newline\urlprefix\url{http://dx.doi.org/10.1111/j.1475-3995.2009.00696.x}

\bibitem{hochstattler2010catbox}
W.~Hochst{\"a}ttler, A.~Schliep,
  \href{http://www.springer.com/new+%26+forthcoming+titles+%28default%29/book/978-3-540-14887-6}{CATBox:
  An Interactive Course in Combinatorial Optimization}, 1st Edition, Springer,
  2010.
\newblock \href {http://dx.doi.org/10.1007/978-3-642-03822-8}
  {\path{doi:10.1007/978-3-642-03822-8}}.
\newline\urlprefix\url{http://www.springer.com/new+%26+forthcoming+titles+%28default%29/book/978-3-540-14887-6}

\bibitem{boyer2004cutting}
J.~Boyer, W.~Myrvold, On the cutting edge: Simplified {O(n)} planarity by edge
  addition, Journal of Graph Algorithms and Applications 8~(3) (2004) 241--273.

\end{thebibliography}
